\documentclass[preprint,amsmath,amssymb,nofootinbib,eqsecnum,tighten]{revtex4}
\usepackage{graphicx}% Include figure files
\usepackage{dcolumn} % Align table columns on decimal point
\usepackage{bm, bbm}      % bold math
\usepackage{curves}
\usepackage{epic}
\usepackage{wasysym}
\usepackage{epsf, times}
\usepackage{subfigure}

\begin{document}

%\preprint{APS/123-QED}

\title{
Dynamics of single polymers under extreme confinement}

\author{
Armin Rahmani$^1$, Claudio Castelnovo$^{1,2}$, Jeremy Schmit$^3$ and
Claudio Chamon$^1$
       }

\affiliation{ $^1$ Department of Physics, Boston University, Boston,
MA 02215 USA,}

\affiliation{ $^2$ Rudolf Peierls Centre for Theoretical Physics,
University of Oxford, UK,}

\affiliation{ $^3$ Department of Physics, Brandeis University,
Waltham MA 02454 USA.}

\date{\today}

\begin{abstract}
We study the dynamics of a single chain polymer confined to a two
dimensional cell. We introduce a kinetically constrained lattice gas
model that preserves the connectivity of the chain, and we use this
kinetically constrained model to study the dynamics of the polymer
at varying densities through Monte Carlo simulations. Even at
densities close to the fully-packed configuration, we find that the
monomers comprising the chain manage to diffuse around the box with
a root mean square displacement of the order of the box dimensions
over time scales for which the overall geometry of the polymer is,
nevertheless, largely preserved. To capture this shape persistence,
we define the local tangent field and study the two-time
tangent-tangent correlation function, which exhibits a glass-like
behavior. In both closed and open chains, we observe reptational
motion and reshaping through local fingering events which entail global
monomer displacement.
\end{abstract}

\maketitle
%
%
%%%%%%%%%%%%%%%%%%%%%%%%%%%%%%%%%%%%%%%%%%%%%%%%%%%%%%%%%%%%%%%%%%%%%%%%%%%%%%

\section{\label{sec: Introduction}
Introduction
    }
In this paper we consider the situation of a single chain polymer
confined within a space smaller than its radius of gyration.  Such a
situation is encountered within the nucleus of a cell where one or
more chromosomes with a radius of gyration on the order of 10 ${\rm
\mu m}$ are confined by the nuclear membrane to a space of order 1
${\rm \mu m}$.  Even in the case of organisms where the genome is
composed of many chromosomes, the situation is distinct from that of
a polymer melt as each chromosome is effectively confined to a
separate sub-volume of the nucleus~\cite{Cremer:01}. Since many
biological processes such as gene suppression and activation require
a rearrangement of the DNA polymer, understanding the dynamics of
confined polymers may yield insight to the dynamics of these
cellular activities. Strongly confined polymers may also be
encountered in the ``lab on a chip'' applications promised by
microfluidic technology~\cite{Whitesides:06}. In these applications
the reaction vessels are $\sim 10 {\rm \mu m}$ microdroplets.

While the equilibrium properties of confined polymers may be
understood based on scaling arguments~\cite{deGennes:79}, the
dynamics of confined polymers are less well understood, and there
has been growing interest in the problem. For instance, the
transport of polymers in confined geometries has been studied in a
variety of contexts, including translocation through
pores~\cite{Muthukumar:01,Bezrukov:96}, diffusion through
networks~\cite{Nykypanchuk:02} and tubes~\cite{Kalb-Chakraborty},
and the packing of DNA within viral capsids~\cite{Spakowitz:05}.
%Here we consider the dynamics of a single
%polymer confined within a box much smaller than the polymer's natural
%radius of gyration.
%
For these highly confined polymers the density profile strongly
resembles that of a polymer melt.  We might naively expect that
since a given section of the polymer interacts primarily with
segments that are greatly separated along the chain, each
segment may be treated as a sub-chain embedded within a melt.
However, this picture is troublesome for dynamical quantities as
reptation theory says that the dynamics is governed by the time it
takes for a given chain to vacate the tube defined by its immediate
neighbors.  With a system consisting of a single polymer, this would
imply that the tube occupies the entire box.  Therefore, we would be
forced to conclude that the chain is completely immobile.  We show
here that reptation-like motion is, in fact, the dominant mode of
deformation of confined polymers. In contrast to the situation in
melts, however, reptational diffusion is not necessarily initiated
by the chain ends, and therefore, cannot be always thought of as
diffusion along a fixed tube.

Polymers confined to thin films have been experimentally shown to have
glassy characteristics~\cite{Keddie:94,Forrest:01}. While this
phenomenon has attracted considerable theoretical attention, it is not
well understood~\cite{deGennes:00,Ngai:98,Jiang:99}. It is also not
known whether glassy behavior occurs in other confined
geometries. Here we point out a connection between lattice polymer
models and Kinetically Constrained Models (KCM) with the chain
connectivity as the analog of the kinetic constraint. Since many KCMs
display glassy behavior at high density it is plausible that polymers
do as well.

In this paper we numerically explore the dynamics of confined
polymers using a kinetically constrained lattice gas model. We find
that monomer diffusion exhibits power law behavior up to densities very
close to the close-packing limit. However, the overall shape of the
chain, as quantified by a tangent-tangent correlation function,
shows a broad plateau at high densities. This apparent paradox is
due to the reptation-like nature of the chain movement. Because the
monomer diffusion is primarily in the direction of the chain
backbone, only relatively small rearrangements of the backbone are
required for the monomers to move distances comparable to the system
size.

The outline of the paper is as follows. In Section~\ref{sec:
particle model} we define our model and employ Monte Carlo
simulations to show that this model reproduces known results for the
dynamic and static properties of unconfined polymers in two
dimensions. In Section~\ref{sec: MSD} we show that the individual
monomers diffuse with a power law in time behavior up to the
close-packing density. In Section~\ref{sec: tantan} we define the
tangent-tangent correlation function and use it to show that the
overall shape of the chain is essentially frozen within the time
scale required by a monomer to diffuse across distances much larger
than the inter-particle seperation. In Section~\ref{sec: tan_dis} we
use a tangent-displacement correlation function to show that the
discrepancy between the monomer diffusion and reshaping time scales
is due to reptation-like diffusion of the polymer along the chain
backbone. Finally, in Section~\ref{sec: conclusions} we summarize
our conclusions.
%
%
%%%%%%%%%%%%%%%%%%%%%%%%%%%%%%%%%%%%%%%%%%%%%%%%%%%%%%%%%%%%%%%%%%%%%%%%%%%%%%

\section{\label{sec: particle model}
A kinetically constrained Lattice gas model
        }
Inspired by the bond fluctuation model~\cite{Kremer:88} of polymer
dynamics and kinetically constrained models (KCM)~\cite{ritort:03}
such as the Kob-Andersen model~\cite{Kob:93,Toninelli:04}, we
propose a KCM for the dynamics of a self-avoiding polymer. The fact
that the monomers constitute a polymer requires the connectivity to
be preserved. Namely, connected (unconnected) monomers must remain
connected (unconnected) during the polymer motion. We begin by
introducing the model in two dimensions with monomers living on the
sites of a square lattice for simplicity. Let us define the polymer
connectivity in the following way. Consider a square box of linear
size $2r$ whose center lies on a given monomer. Any other monomer
that lies inside or on the boundary of this box is defined as a
box-neighbor of the monomer at the center. Clearly, if monomer A is
a box-neighbor of monomer B, then monomer B is also a box-neighbor
of monomer A. We define two monomers as being connected by a bond if
and only if they are box-neighbors.
A monomer with no box neighbors is an isolated polymer of length one.
A monomer with only one box-neighbor is the end-point of a polymer.
A monomer with two box neighbors is a point in the middle of a polymer
and a monomer with more than two box neighbors corresponds to a branching
point along a polymer. Depending
on the initial monomer positions, multiple open or closed chains can
be modeled. Also by using a d-dimensional hyper-cube instead of a
square, the model can be immediately generalized to higher dimensions.

\begin{figure}[ht]
\vspace{0.6 cm}
\begin{center}
\includegraphics[angle=0,scale=0.7]{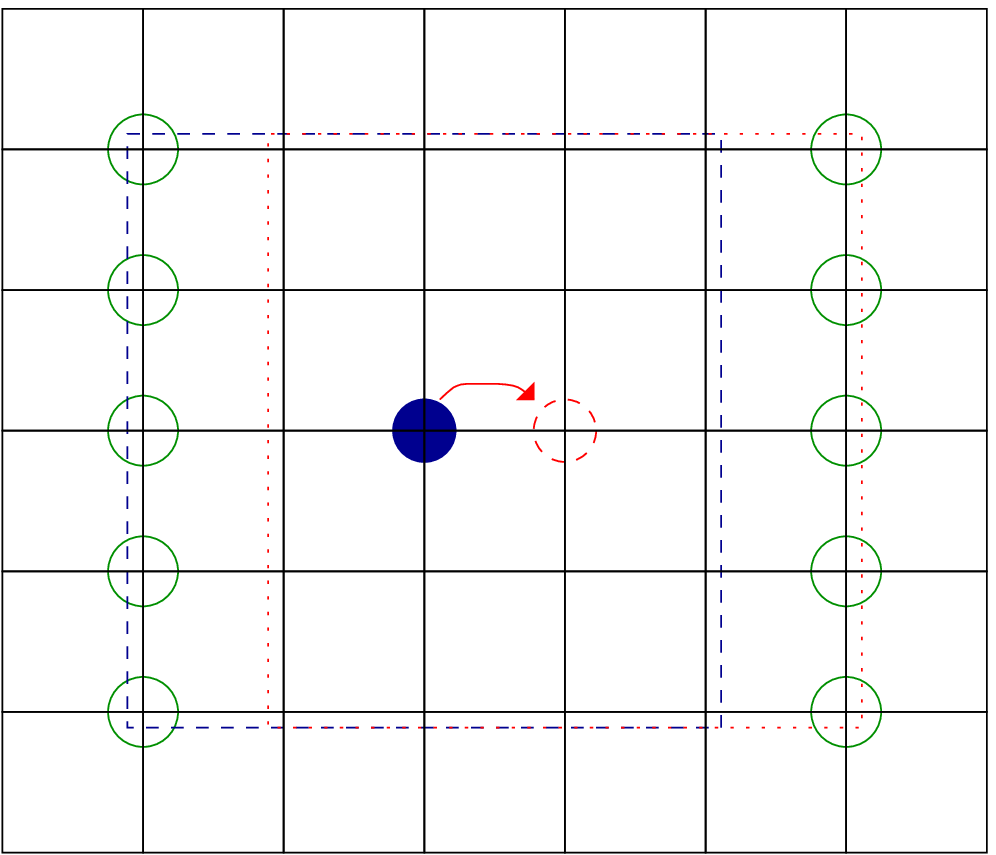}
\end{center}
\caption{ \label{fig:dynamics} For a model with an $r=2$ box, a
monomer can hop in a given direction if the destination site, as well
as the ten sites which the old and new box differ by, are empty.}
\end{figure}

The dynamics is defined as follows. A monomer can hop to a nearest
neighbor unoccupied site if it has exactly the same box-neighbors
before and after the move, as in Fig.~\ref{fig: hard particle
example}. If no monomer enters the box associated with the moving monomer
and no monomer falls out of it during the move, the box will contain the
exact same monomers before and after the move. In other words, all
the $2(2r+1)$ sites ($2(2r+1)^{(d-1)} $ in $d$ dimensions) that
enter or exit the box as it is moved to the new position, must be
unoccupied, as shown in Fig.~\ref{fig:dynamics}. In our model as in
many kinetically constrained lattice gas models~\cite{ritort:03}, we
take the energy to be independent of the configuration, resulting in
constant hopping rates. We assume that the allowed moves take place
at unit rate. So if $n(x,y)$ is defined to assume the value $1$ for
occupied sites and $0$ for empty sites, and
$\bar{n}(x,y)\equiv1-n(x,y)$, the rate of hopping to the right out
of site $(x,y)$  is given by
\begin{equation}\label{rate}
   w_{\rightarrow}(x,y)=n(x,y)\bar{n}(x+1,y)\prod_{j=-r}^{j=+r}\bar{n}(x+r+1,y+j)\bar{n}(x-r,y+j),
\end{equation}
and by similar expressions for the other directions. The dynamics
forbids monomers that are unconnected from getting too close to each
other and therefore ensures self-avoidance. Notice that all the moves are
reversible because, as seen in Fig.~\ref{fig:dynamics}, any particle
that was allowed to hop to a nearest neighbor empty site is allowed
to hop back to its original position.
\begin{figure}[ht]
\vspace{0.6 cm}
\begin{center}
\includegraphics[angle=0,scale=0.7]{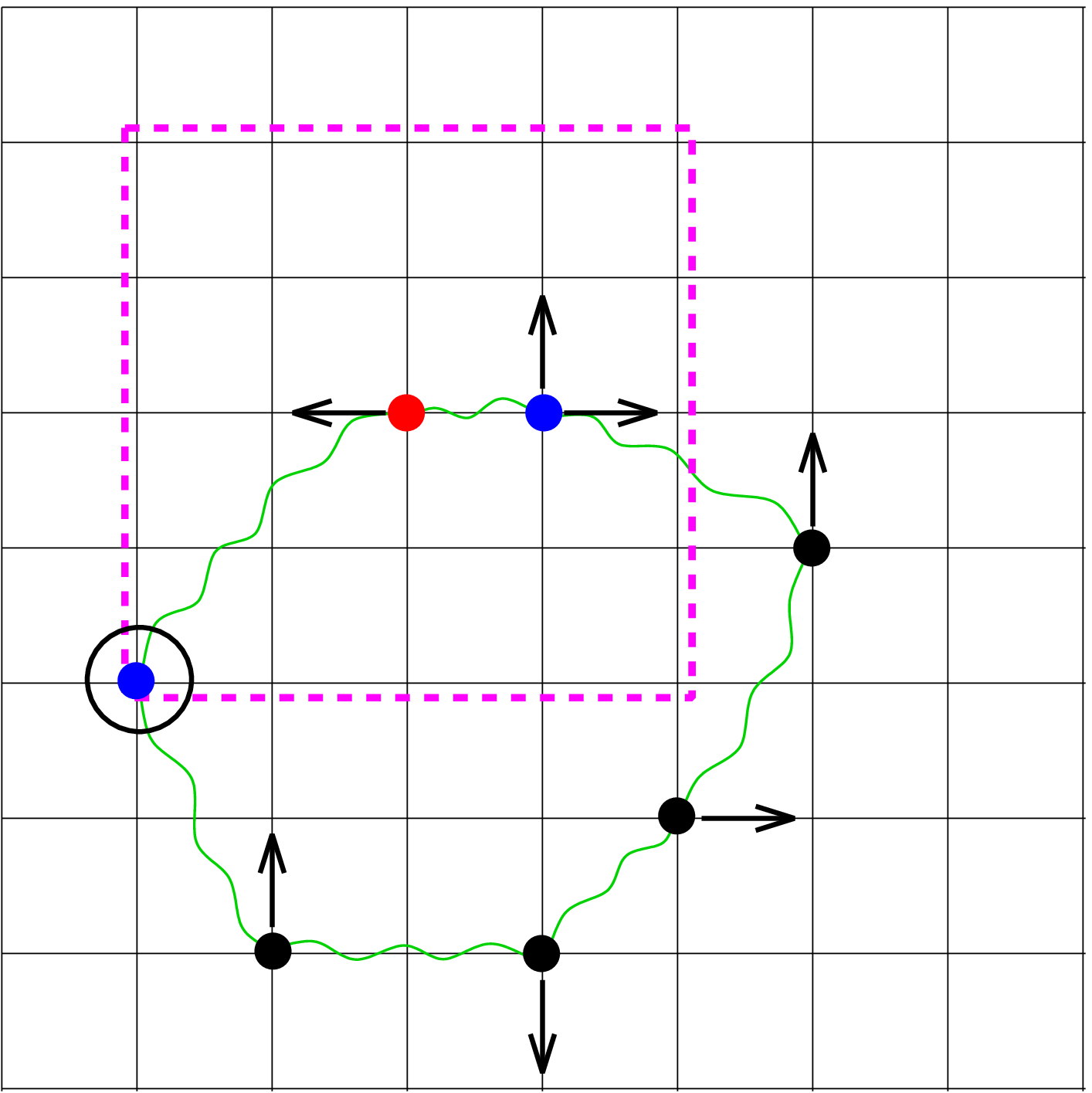}
\end{center}
\caption{ \label{fig: hard particle example} An example of a
configuration that satisfies the initial conditions discussed in the
text to have a single closed polymer.
This is illustrated explicitly for the particle in red: the
box of size $2r$ ($r=2$) is indicated by the dashed purple square,
and the two other particles inside the box are colored in blue. The
same condition holds for all the particles in the system. All the
(nearest-neighbor) particle moves allowed by the kinetic constraint
are shown for each particle with arrows along the corresponding
lattice edge. One particle in this configuration is temporarily
frozen (circled blue particle), and its move is subordinated to the
move of one of the two particles in its box of size $2r$. Notice
that the initial sequence of particles, represented by the wiggly
green line, is clearly preserved by the allowed moves. }
\end{figure}
We choose the smallest value of $r$ for which the model behaves like
a polymer while allowing for shorter simulation times. The $r=1$
case is too restrictive to model different modes of motion. For
example a polymer lying along a straight line is forbidden in the
$r=1$ model to undergo one-dimensional translation. We choose $r=2$
as it is found to adequately describe the free polymer dynamics, as
shown by our numerical simulations.

Monte Carlo (MC) simulations are used to study the model. A
particularly time-efficient algorithm is achieved by storing two
representations of the system at each MC step. One consists of the
position vector of all the monomers, and the other is the
configuration matrix of the lattice, with unoccupied sites having
value zero and occupied sites value one. This allows us to chose a
monomer at random from the position vector (rather than a site at
random from the whole lattice) and quickly determine if the monomer
is allowed to move in a randomly chosen direction by checking the
values of at most eleven elements (the nearest neighbor site plus
the sites by which the old and new box differ) in the configuration
matrix. Note that it is possible to define an alternative
model by using a circle of radius $r=2$ instead of a square box of
side $2r=4$, which would require the same amount of computational
effort because the same number of sites, namely ten sites, would
enter or leave the circle drawn around the moving monomer.

\begin{figure}[ht]
\vspace{0.6 cm}
\begin{center}
\includegraphics[angle=0,scale=0.7]{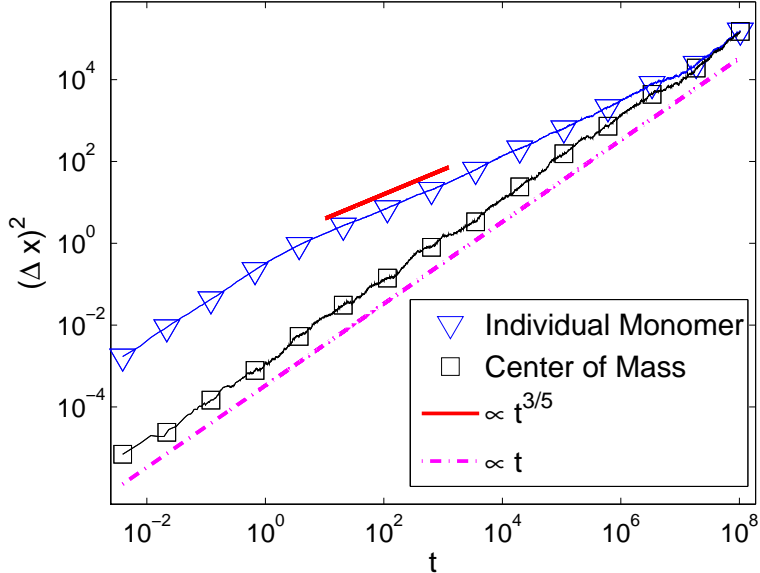}
\end{center}
\caption{ \label{fig:rouse} Center of mass and individual monomer
mean square displacement of a free polymer. The results are shown
for $N=256$. }
\end{figure}

The $r=2$ model with a square box proved to give consistent results
with the ones available in the literature. Namely, the time-averaged
radius of gyration squared of the polymer computed for our model in
an infinite box scales as $R^2\propto N^{1.451\pm0.084}$, which is
consistent with Flory's theoretical result of $R^2\propto
N^{\frac{3}{2}}$~\cite{Flory:49}. The dynamics of the polymer in
unconfined environments is also compatible with the Rouse model to a
good approximation. As shown in Fig.~\ref{fig:rouse}, the mean
square displacement of the center of mass is diffusive with a
diffusion constant that scales as $N^{-0.96}$ compared to the
$N^{-1}$ theoretical value. Throughout the paper, time and length
are measured in units of Monte-Carlo steps and lattice spacings
respectively. Only four values of $N=128,256,512,1024$ were used for
these consistency checks but the fits were very close to the
theoretical predictions. The individual monomer mean square
displacement is diffusive at very short times followed by an
intermediate-time subdiffusive behavior and a cross-over to a final
diffusive behavior at long times as each monomer begins to move with
the center of mass. The subdiffusive MSD can be fit with an exponent
of $0.5968\pm0.0008$ over the two-decade interval of $10^1<t<10^3$
which is consistent with the theoretical value of $\frac{3}{5}$ and
the bond fluctuation results~\cite{Kremer:88}. Because of the
cross-over to diffusive behavior at long times the curve fits well
to a higher exponent of $0.6516\pm0.0008$ over the longer interval
of $10^1<t<10^6$.

\begin{figure}[ht]
\vspace{0.6 cm}
\begin{center}
\includegraphics[angle=0,scale=0.4]{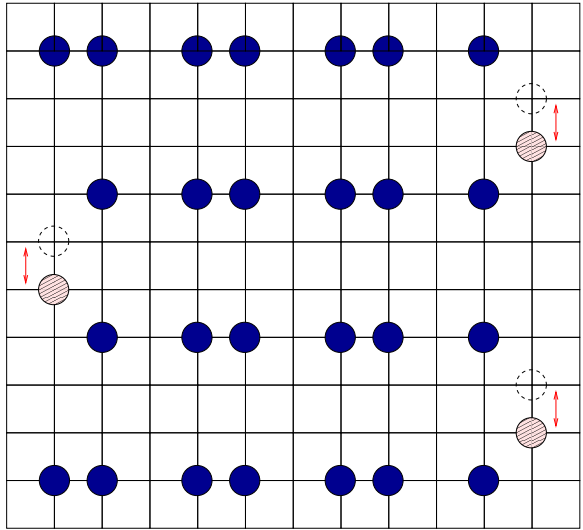}
\end{center}
\caption{ \label{fig:fully packed} A fully packed configuration
of the model described in the text. Notice that only the shaded monomers
are allowed to move at any given time.}
\end{figure}

In the present paper we focus on open and closed polymers without any
branching. The close-packing density of such polymers clearly has an
upper limit. For a single open chain for example, each monomer except
for the two end-points has exactly two box-neighbors. The maximally
packed configuration is achieved once the distance along the chain
between the consecutive monomers alternates between one and two
lattice spacings, and the distance between parallel segments of the
folded polymer equals $3$ lattice spacings, as depicted in
Fig.~\ref{fig:fully packed}. If we have $N$ monomers on an
$L\times L$ lattice with $(L+1)^2$ sites, the fully-packed
configuration attains a thermodynamic-limit density
$\rho=\frac{N}{(L+1)^2} \to \frac{2}{9}$ ($\frac{2}{3^d}$ in $d$
dimensions). Note that except for fluctuations at the U-turns, the
polymer is completely frozen at close-packing.
%
%
%
%
%%%%%%%%%%%%%%%%%%%%%%%%%%%%%%%%%%%%%%%%%%%%%%%%%%%%%%%%%%%%%%%%%%%%%%%%%%%%%%

\section{\label{sec: MSD}
Mean square monomer displacement
        }
\begin{figure}[ht]
\vspace{0.6 cm}
\begin{center}
\includegraphics[angle=0,scale=0.8]{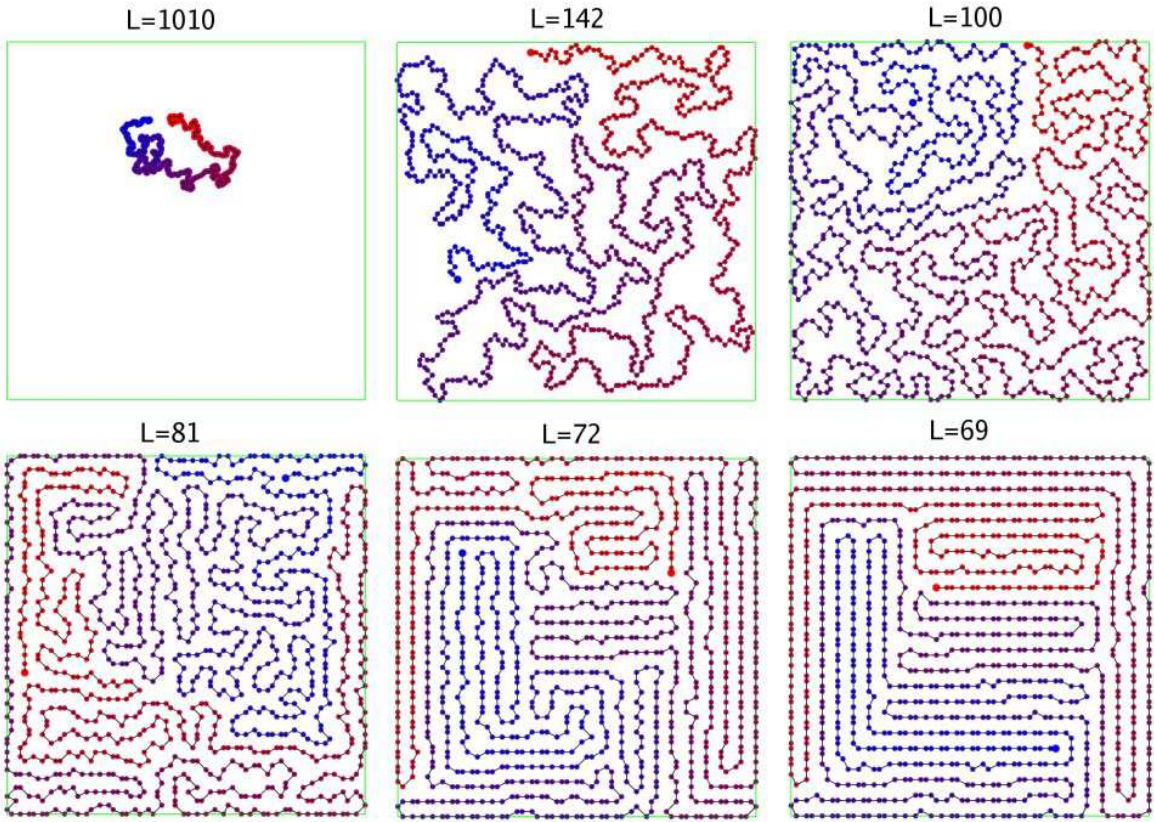}
\end{center}
\caption{\label{fig:shrink}
Snapshots of a polymer with $N=1024$ monomers.
As the box shrinks, the polymer gets more and more confined.
At densities close to full-packing, a geometry resembling
that of a compact (Hamiltonian walk) polymer is formed.
}
\end{figure}

The question of whether or not placing a self-avoiding polymer in a
highly confining environment can freeze its motion can be addressed
by measuring the statistical average of the mean square
displacement as a function of time
\begin{equation}
\label{eq: mean_square-displacement}
c(t)
=
\langle
  \frac{1}{N}
    \sum_{i=1}^{i=N}
      \left[\vphantom{\sum}
        \vec{x_i}(t+t_w)-\vec{x_i}(t)
      \right]^2
\rangle,
\end{equation}
where $\vec{x_i}$ is the position of the $i$-th monomer and $t_w$ is
the waiting time between the preparation of the sample and the
measurement. Throughout the paper, we denote the ensemble average by
$\langle\ldots\rangle$. We prepare random samples with different
densities by placing the polymer in a large box and gradually
reducing the size of the box. This is achieved by forbidding the
monomers to move to the edges of the box, which corresponds to an
infinite repulsive potential at the boundary, and removing one
vertical and one horizontal edge line once they have become
completely empty. After each shrinking process, the system is
confined to a smaller square box. Shrinking and measurements are done
in series. Specifically, we start from a very low density of
$\rho=0.00010$ and after each shrinking step we let the system run
for $1000$ Monte-Carlo steps before trying to shrink further.
%in the
%attempt to reduce possible memory effects induced by the specific
%details of the chosen shrinking procedure.
For measurements involving a long waiting time ($t_w=10^7$ steps),
i.e., for densities $\rho\simeq0.0010,0.050,0.10,0.15$ and $\rho>0.195$,
subsequent shrinking steps are preformed starting from the
post-measurement configurations.
%because the possible memory effects
%are expected to persist longer at higher densities and the long
%measurement time could help reduce them further.
With this method we are able to reach densities of approximately
$\rho=0.21$, compared to the limiting theoretical value of $\rho =
2/9 \simeq0.222$. At high densities, as shown in
Fig.~\ref{fig:shrink}, the overall geometry of the polymer resembles
that of a compact polymer described by a Hamiltonian path, i.e., a
path which visits all sites exactly once, exploring a lattice with
lattice spacing three times larger than in the original one. At
these high densities our model resembles a semi-flexible polymer
because the chain is able to attain a higher packing density in the
direction parallel to the backbone (average monomer spacing equal to
$\frac{3}{2}$ lattice spacings) than in the direction perpendicular
to the backbone (average monomer spacing equal to $3$ lattice
spacings) (see Fig. \ref{fig:fully packed}).

\begin{figure}[ht]
\vspace{0.6 cm}
\begin{center}
\includegraphics[angle=0,scale=0.7]{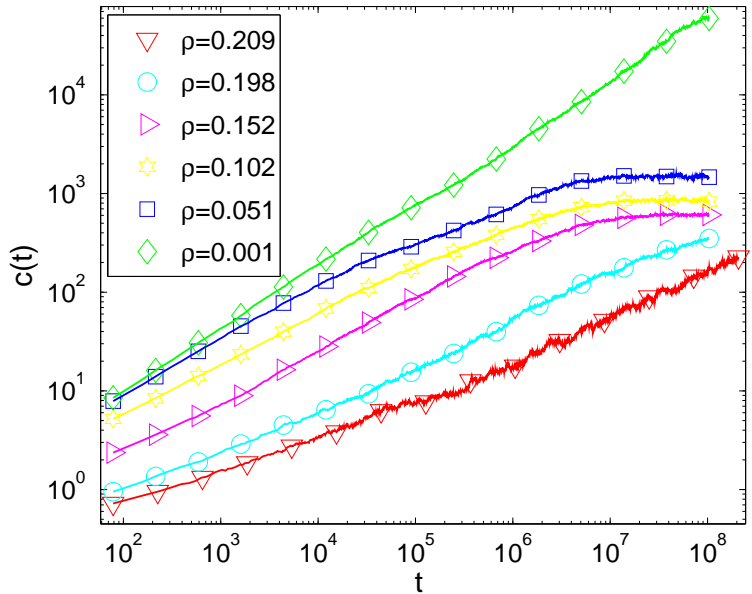}
\end{center}
\caption{ \label{fig:MSD_tau} Mean square displacement for $N=256$ and
different box sizes (i.e., different densities) as a function of time.
The $\rho=0.001$ curve can be fit with a $0.63$ exponent which
is close to the intermediate-time regime of the Rouse model. The
results for other values of N are qualitatively very similar. }
\end{figure}

Measurements are done with two values of waiting time, $t_w=10^7$
and $t_w=1.1\times10^8$ Monte-Carlo steps over a period of $t=10^8$
steps. (For the highest density we used $t=2\times10^8$
instead.) The mean square displacement shows time translation
invariance up to the highest densities achieved. We study the
behavior of $c(t)$ for $N=128,256,512,1024$ at the densities listed
above. The root mean square displacement $\sqrt{c(t)}$ is a measure
of how much the monomers have moved. As shown in
Fig.~\ref{fig:MSD_tau}, $c(t)$ increases with power law behavior and
finally saturates with a limiting root mean square displacement of
the order of the box size. For very large box sizes (lowest density)
as well as very small box sizes (very high densities), the
saturation plateau is not always reached within the measurement
time. However, the maximum value of $\sqrt{c(t)}$ is still of the
same order of the box dimensions. Although the dynamics slows down
at high confinement, each monomer manages to move an average
distance comparable to the box size over our measurement time $t$.
Visually observing the polymer motion, however, clearly shows a more
complex scenario where at high densities the overall geometry of the
polymer is largely preserved (see Fig.~\ref{fig:shrink}). Indeed, we
will see that the tangent-tangent correlation function, although
time-translation invariant at low and intermediate densities,
exhibits signs of aging at higher densities. In the following
sections we discuss in detail the shape persistence as well as the
mechanisms by which the polymer shape changes as the box size is
reduced.
%
%
%
%
%%%%%%%%%%%%%%%%%%%%%%%%%%%%%%%%%%%%%%%%%%%%%%%%%%%%%%%%%%%%%%%%%%%%%%%%%%%%%%

\section{\label{sec: tantan}
Tangent field Correlation
        }
\begin{figure}[ht]
\vspace{0.6 cm}
\begin{center}
\includegraphics[angle=0,scale=0.4]{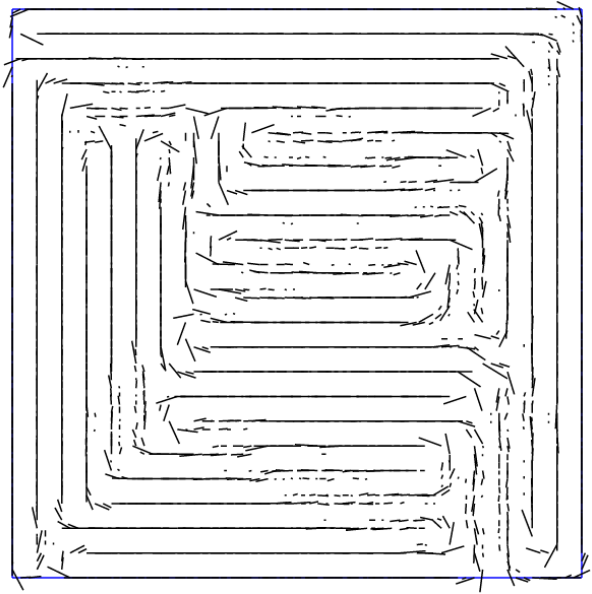}
\end{center}
\caption{ \label{fig:tang_ex} An example of the coarse-grained
tangent field $\vec{s}_{\vec{x}}(t)$ }
\end{figure}

The large values reached by $\sqrt{c(t)}$ within our simulation
times even for very high densities indicate that confinement does
not freeze the motion of the monomers. The overall shape or geometry
of the polymer, however, exhibits global persistence at high
densities as observed via direct visualization of the dynamics. In
order to systematically study shape persistence and reshaping, we
introduce the concept of a tangent field, a vector field defined on
the entire lattice which captures the overall shape of the polymer.
We define the instantaneous tangent field as
\begin{equation}
\label{eq: tan}
\vec{s}_{\textrm{inst}}(\vec{x},t)
=
\left\{
  \begin{array}{ll}
    \vec{x}_{i+1}(t) - \vec{x}_{i-1}(t)
    &
    \textrm{if $\vec{x} = \vec{x}_{i}(t)$}
    \\
    \vec{0}
    &
    \textrm{otherwise},
  \end{array}
\right.
\end{equation}
$\vec{x}_{i}(t)$ being the position of monomer $i$ at time $t$.
The tangent field is defined in a symmetric way so that labeling the
monomers in reverse
order only changes the direction of the field. In an open chain, the
definition needs to be modified at the end points. Note that the
tangent field is indexed by a position in space and not by a monomer
number; this allows us to compare the shapes at different times, independently of the monomer motion. Since local vibrations of the polymer
do not change the overall geometry, we seek a quantity that is
insensitive to these vibrations. Coarse-graining the field by
time averaging over a carefully chosen interval removes the local
vibrations and results in a smeared field $\vec{s}_{\vec{x}}(t)$
which captures the overall geometry, as shown in
Fig.~\ref{fig:tang_ex}. We have chosen the time
interval to be $75$ Monte Carlo steps (or $75 \times N$ single monomer
attempts) which is sufficient to allow for several vibrations. Since
the success rate of the Monte Carlo attempts at high density is
found to be around $1/10$, this interval corresponds to
roughly $7$ moves per monomer.
In terms of the coarse-grained tangent field defined above,
we define the tangent-tangent correlation function
\begin{equation}
\label{eq: tantan}
c_s(t,t_w)
=
\langle
  \frac{1}{N}
  \sum_{\textrm{all}\,\vec{x}}
    \left[\vphantom{\sum}
      \vec{s}_{\vec{x}}(t+t_w)
      \cdot
      \vec{s}_{\vec{x}}(t_w)
    \right]
\rangle
\end{equation}
as a measure
of the overlap of the tangent field at times $t+t_w$ and $t_w$.

\begin{figure}[ht]
\vspace{0.6 cm}
\begin{center}
\includegraphics[angle=0,scale=0.7]{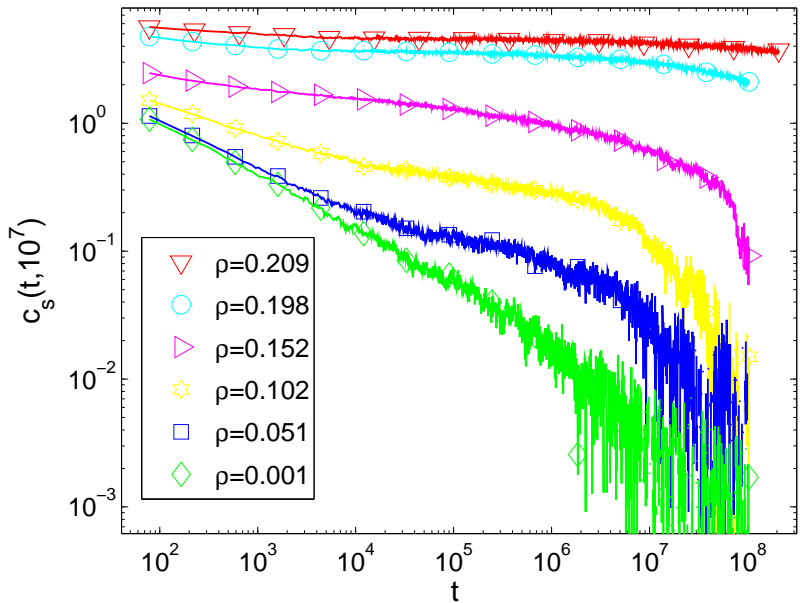}
\end{center}
\caption{ \label{fig:tantan_tau} Tangent-tangent correlation
function for $256$ monomers and different box sizes. The $\rho=0.001$
curve fits well to a power law of exponent $0.42$. Note that the
correlation functions are not normalized. }
\end{figure}

\begin{figure}

\begin{center}
%\subfigure[Low density]
{\includegraphics[angle=0,scale=0.6]{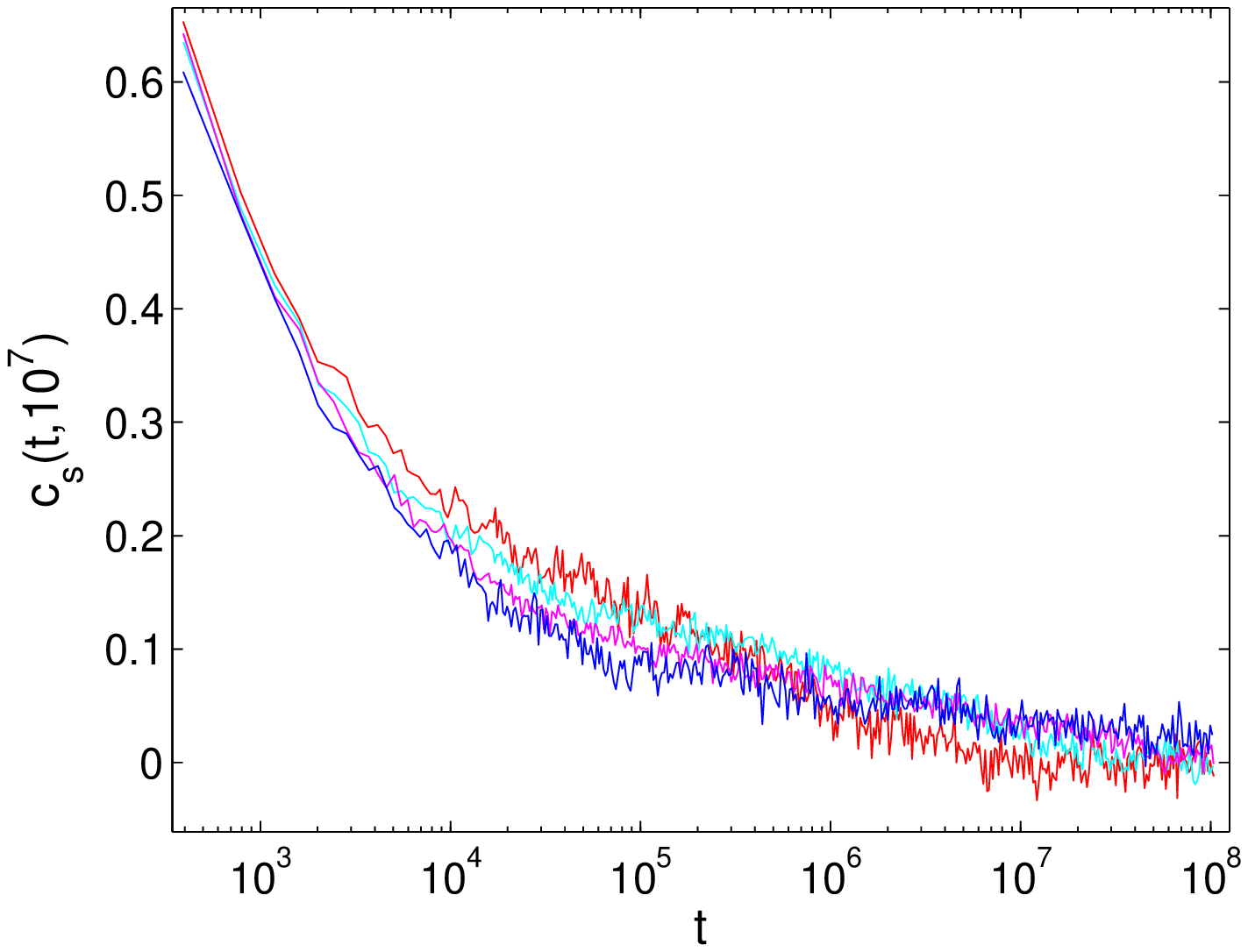}}
%\hspace {1in} \subfigure[High density]
{\includegraphics[angle=0,scale=0.6]{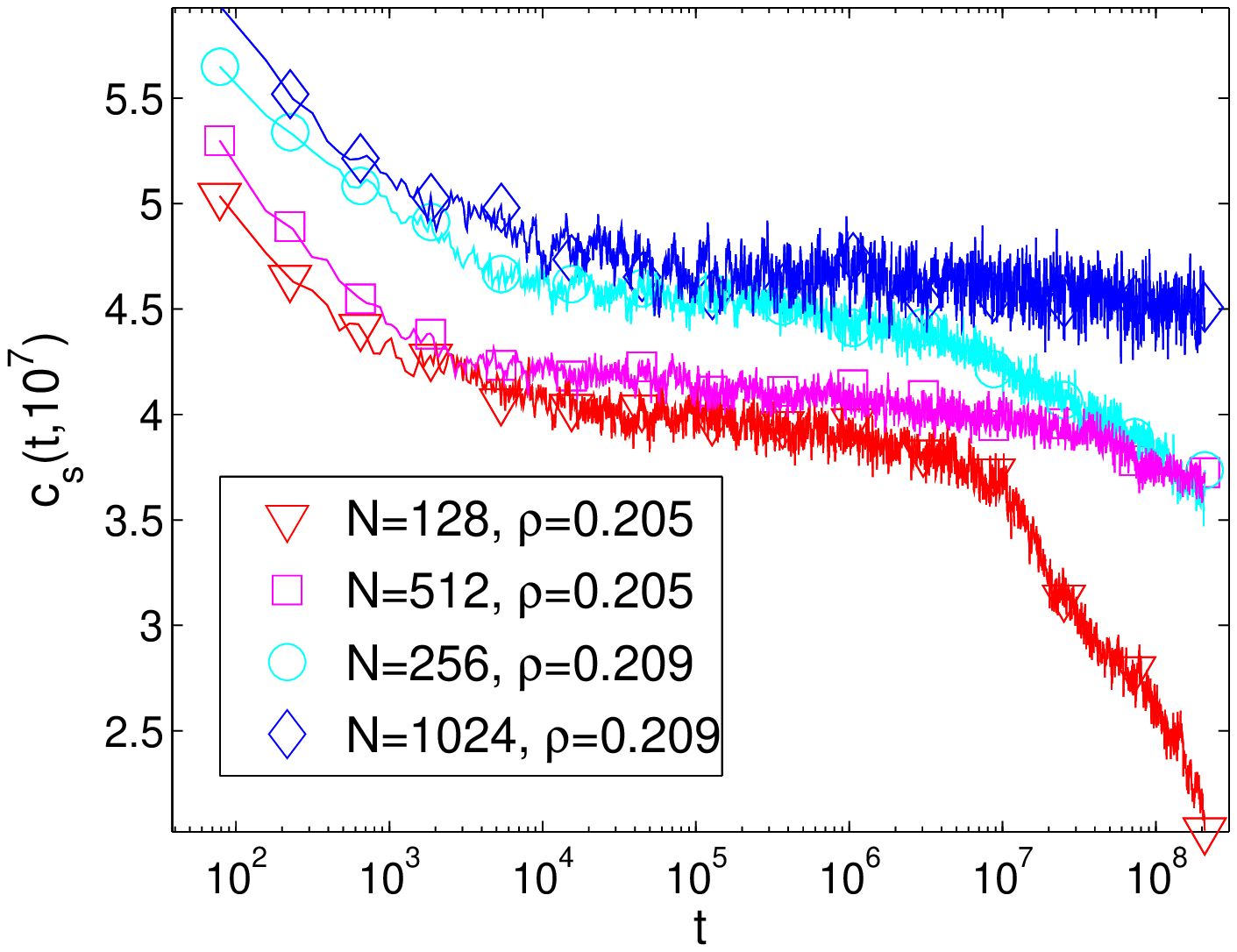}}

\end{center}
\caption{
\label{fig:tantan_size}
The tangent-tangent correlation function for four different polymer sizes.
Top: At low density $\rho=0.05$, the curves are independent of $N$.
Bottom: At the highest achieved density $\rho \simeq 0.20-0.21$,
the width of the emergent plateau increases with $N$
}
\end{figure}

As shown in Fig.~\ref{fig:tantan_tau}, $c_s(t,t_w)$ decays as a
power low in $t$ for very low densities. As the density increases, a
second time scale emerges and at the highest densities we clearly
see an initial decay followed by a broad plateau and a secondary
decay. (Notice the use of a logarithmic scale on both axes.) The
time-averaging of the tangent field hides the fast mode responsible
for the initial decay and causes the correlation to have a smaller
initial value at lower densities. The correlation function~(\ref{eq:
tantan}) does not depend on the value of $N$ at low densities, as
seen in Fig.~\ref{fig:tantan_size}, while at higher densities we
observe broader plateaux and longer decorrelation times as the
number of monomers $N$ is increased.

\begin{figure}

\begin{center}
%\subfigure[Low density]
{\includegraphics[angle=0,scale=0.6]{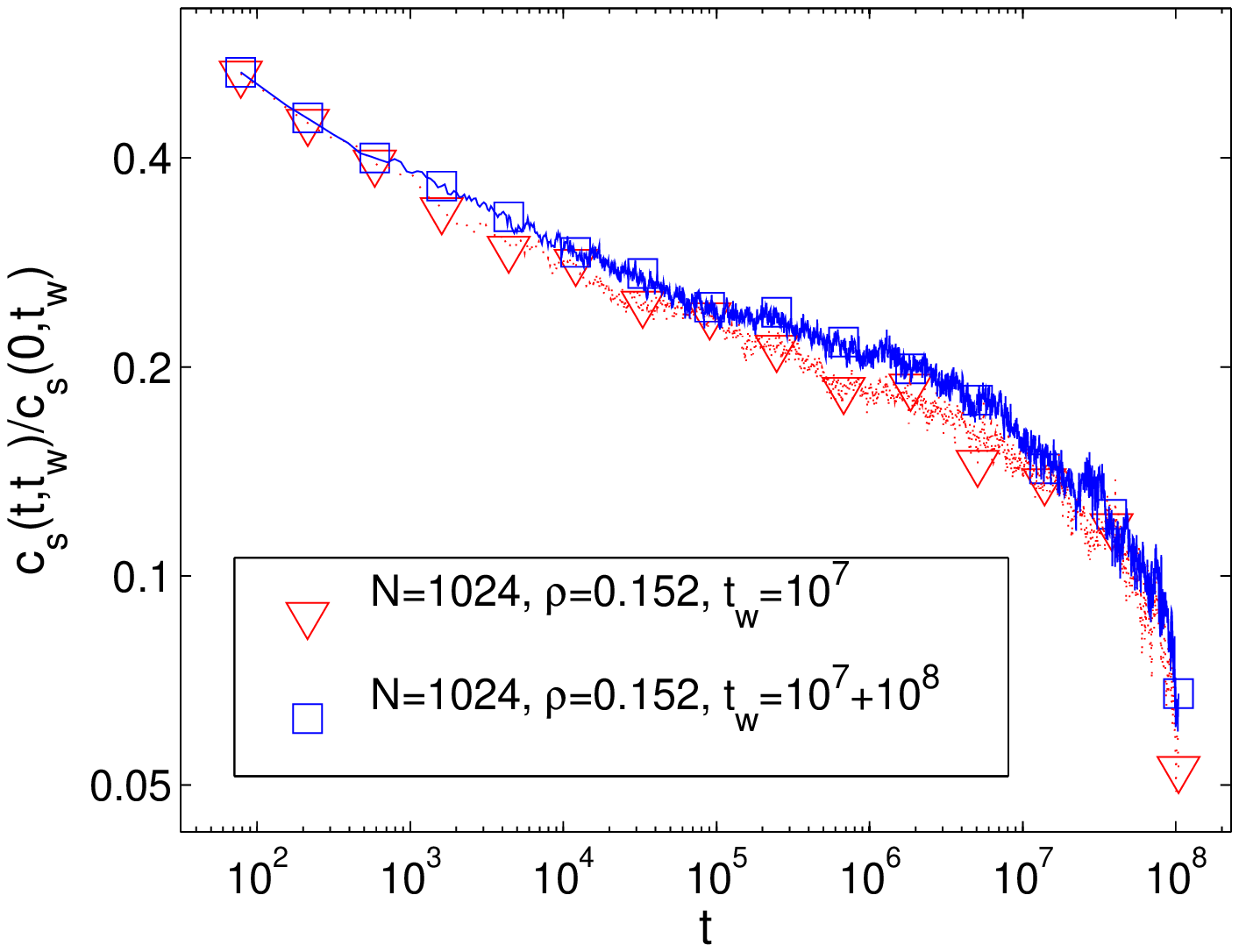}}
%\hspace {1in} \subfigure[High density]
{\includegraphics[angle=0,scale=0.6]{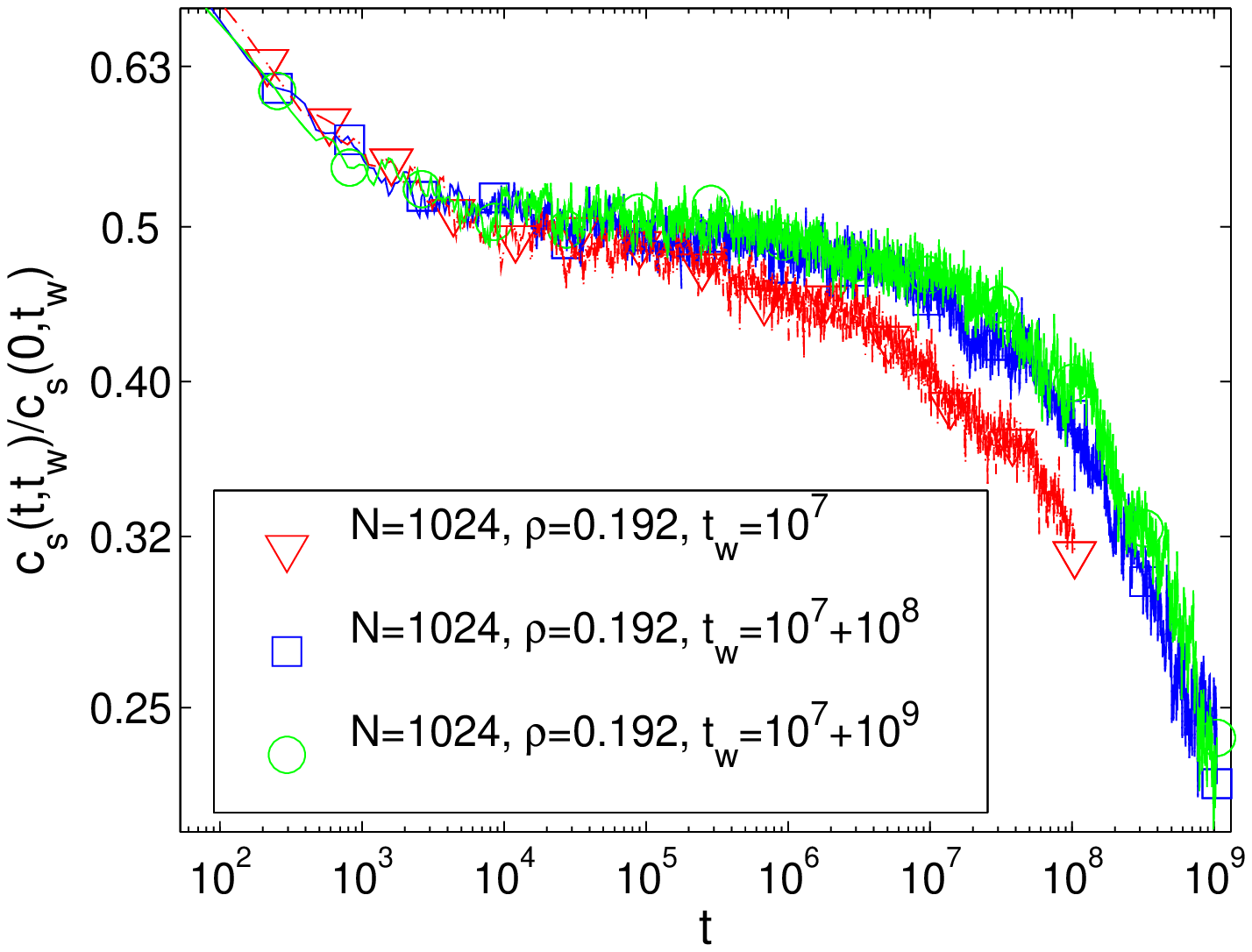}}

\end{center}
\caption{ \label{fig:aging} The normalized tangent-tangent
correlation function $c_s(t,t_w) / c_s(0,t_w)$ vs. time for different waiting times shown in log-log plots.
Top: Up to intermediate densities, before the appearance of a clear
plateau, time translation invariance is not broken. Bottom: At
higher densities, where a broad plateau has emerged, we observe that
the second decay occurs at a longer time scale. With increasing the waiting times the correlation function approaches equilibrium but there is still a systematic shift between $t_w\approx10^9$ and $t_w\approx10^8$ curves indicating slower decay for the older system. }
\end{figure}

To check for time-translation invariance, we ran the simulations two more times after increasing the waiting time $t_w$ by an order of
magnitude each time. The mean square displacement exhibits time-translation
invariance at all densities. For the tangent-tangent correlation
function, however, time-translation invariance is respected at low
densities but violated at the highest densities where a broad
plateau emerges. It appears that at high densities the average
distance between monomers slowly evolves with time, so that the
initial value of the tangent-tangent correlation function
$c_s(0,t_w)$ depends on $t_w$. If we normalize the correlation
function using its value at the beginning of the measurement and plot
$c_s(t,t_w) / c_s(0,t_w)$ as a function of time, the violation
of time-translation invariance suggests the existence of aging
effects, a comprehensive study of which is beyond the scope of the
present paper. For $t_w=1.01\times10^9$, the system has almost equilibrated but there is still a systematic shift toward longer times compared with the $t_w=1.1\times10^8$ correlation function. Fig.~\ref{fig:aging} summarizes the above
observations. Moreover, Fig.~\ref{fig:tantan_size} suggests longer equilibration times for larger systems and in the thermodynamic limit ($N\rightarrow\infty$) the aging effects are expected to survive for arbitrarily large $t_w$.
%
%
%
%
%%%%%%%%%%%%%%%%%%%%%%%%%%%%%%%%%%%%%%%%%%%%%%%%%%%%%%%%%%%%%%%%%%%%%%%%%%%%%

\section{\label{sec: tan_dis}
Tangent-Displacement correlation
        }
\begin{figure}[ht]
\vspace{0.6 cm}
\begin{center}
\includegraphics[angle=0,scale=0.7]{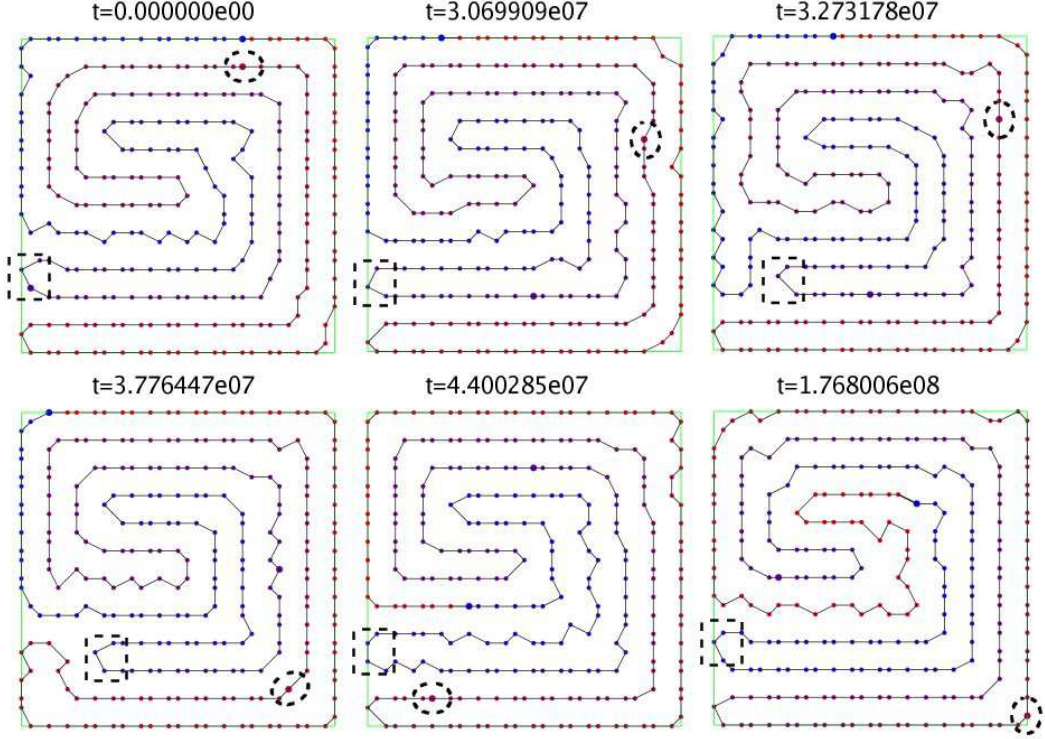}
\end{center}
\caption{ \label{fig:reptation_loop} Snapshots of a realization with
$256$ monomers arranged in a single closed polymer chain,
at density $\rho=0.209$. Pure reptation and fingering
events are visibly responsible for monomer motion. The dashed rectangle
highlights the tip of a finger taking part in a fingering event. The dashed
ellipse shows the reptational motion of a tagged monomer.
}
\end{figure}

As seen in the previous sections, confinement slows down the motion
of individual monomers in a polymer chain, but it does not
fundamentally change the characteristics of their mean square
displacement. It does, however, have a profound effect on reshaping.
Without any reshaping, the only possible motion can happen via
reptation, i.e., when the monomers move back and forth along a fixed
path. With the exception of the unlikely event of the two end-points
finding each other, reptation without reshaping is not possible in
open chains. Indeed, by studying closed loops in detail we do find
that longitudinal diffusion is the main mechanism for motion at high
densities. The existence of large root mean square displacements in
strongly confined open chains and in the absence of major reshaping
can be explained by noting that local reshaping events with only a
minor contribution to the tangent-tangent decorrelation allow for
global monomer motion through a reptation-like process.
Fig.~\ref{fig:reptation} shows one instance of such behavior in a
particularly mobile realization. The mechanisms shown in
Fig.~\ref{fig:reptation}, namely end-point initiated reptation and
"fingering" events, are observed in other realizations as well. A
finger is formed when the chain makes a 180-degree bend resulting in
two adjacent segments of the polymer running antiparallel to each
other. A fingering event occurs when a finger retracts making room
for the extension of another finger. Even in the case of closed
loops where pure reptation is possible, reptation is usually
accompanied by local fingering events as shown in
Fig.~\ref{fig:reptation_loop}. To explicitly quantify the
contribution of reptation to highly confined motion, we define
tangent-displacement and normal-displacement correlation functions
as follows:
\begin{figure}[ht]
\vspace{0.6 cm}
\begin{center}
\includegraphics[angle=0,scale=0.7]{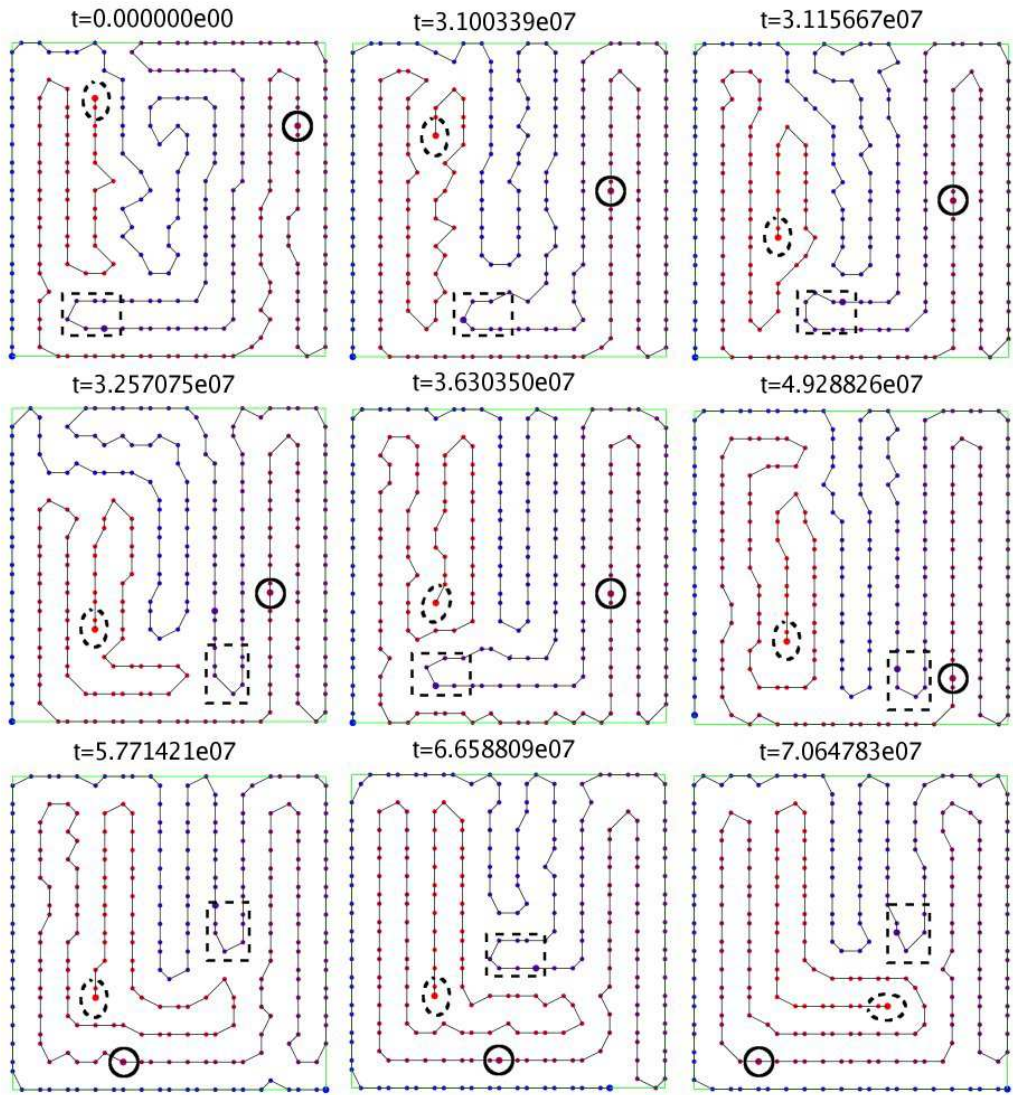}
\end{center}
\caption{ \label{fig:reptation} Snapshots of a particularly mobile
realization of a $256$-monomer chain at density $\rho=0.209$.
Fingering events are observed as well of end-point reptation
accompanied by local rearrangements. An end-point initiated reptation is highlighted
with a dashed ellipse at the moving end-point. The dashed rectangle shows the tip of a finger
which participates in a fingering event during which the tagged monomer shown with a solid circle, for example, moves through reptation.}
\end{figure}
\begin{eqnarray}
\label{eq: tandis}
c_t(t)
&=&
\langle
  \frac{1}{N}
  \sum_{i=1}^{N}
    \left\{\vphantom{\sum}
       \frac{\vec{s}_{\vec{x}_i}(t_w) }{|\vec{s}_{\vec{x}_i}(t_w)|}
       \cdot
       \left[\vphantom{\sum}
         \vec{x}_{i}(t_w+t)
     -
     \vec{x}_{i}(t_w)
       \right]
    \right\}^2
\rangle
\\
\label{eq: nordis}
c_n(t)
&=&
\langle
  \frac{1}{N}
  \sum_{i=1}^{N}
    \left\{\vphantom{\sum}
      \frac{\vec{n}_{\vec{x}_i}(t_w)}{|\vec{n}_{\vec{x}_i}(t_w)|}
      \cdot
      \left[\vphantom{\sum}
        \vec{x}_{i}(t_w+t)
    -
    \vec{x}_{i}(t_w)
      \right]
    \right\}^2
\rangle,
\end{eqnarray}
where $\vec{n}$ is the normal field defined as
$\vec{n}_{\vec{x}}(t)=\hat{z}\times\vec{s}_{\vec{x}}(t)$ and the polymer
-- and therefore $\vec{s}_{\vec{x}}(t)$ -- belongs to the $xy$ plane.
By comparison with Eq.~(\ref{eq: mean_square-displacement}), one can show
that
\begin{equation}
\label{eq: decomp}
    c(t)=c_t(t)+c_n(t).
\end{equation}
Therefore, the correlation functions~(\ref{eq: nordis})
and~(\ref{eq: tandis}) are the contributions to the mean square
displacement due to the transverse and longitudinal motion relative
to the initial polymer orientation. At very high densities and up to
time scales smaller than the beginning of the secondary decay in the
tangent-tangent correlation function Eq.~(\ref{eq: tantan}), $c_t$
is a good measure of the reptational contribution to the confined
motion. This is due to the fact that over such time scales the
polymer shape is largely preserved. Additionally, over the same time
scales the mean square displacement is smaller than the average chain
length between major bends, which as seen in
Fig.~\ref{fig:reptation} is a large fraction of the box size. If the
polymer undergoes major reshaping or the monomers move through the
bends of the folded polymer, reptation would no longer be equivalent
to the motion along the original orientation.

\begin{figure}

\begin{center}
%\subfigure[Low density]
{\includegraphics[angle=0,scale=0.6]{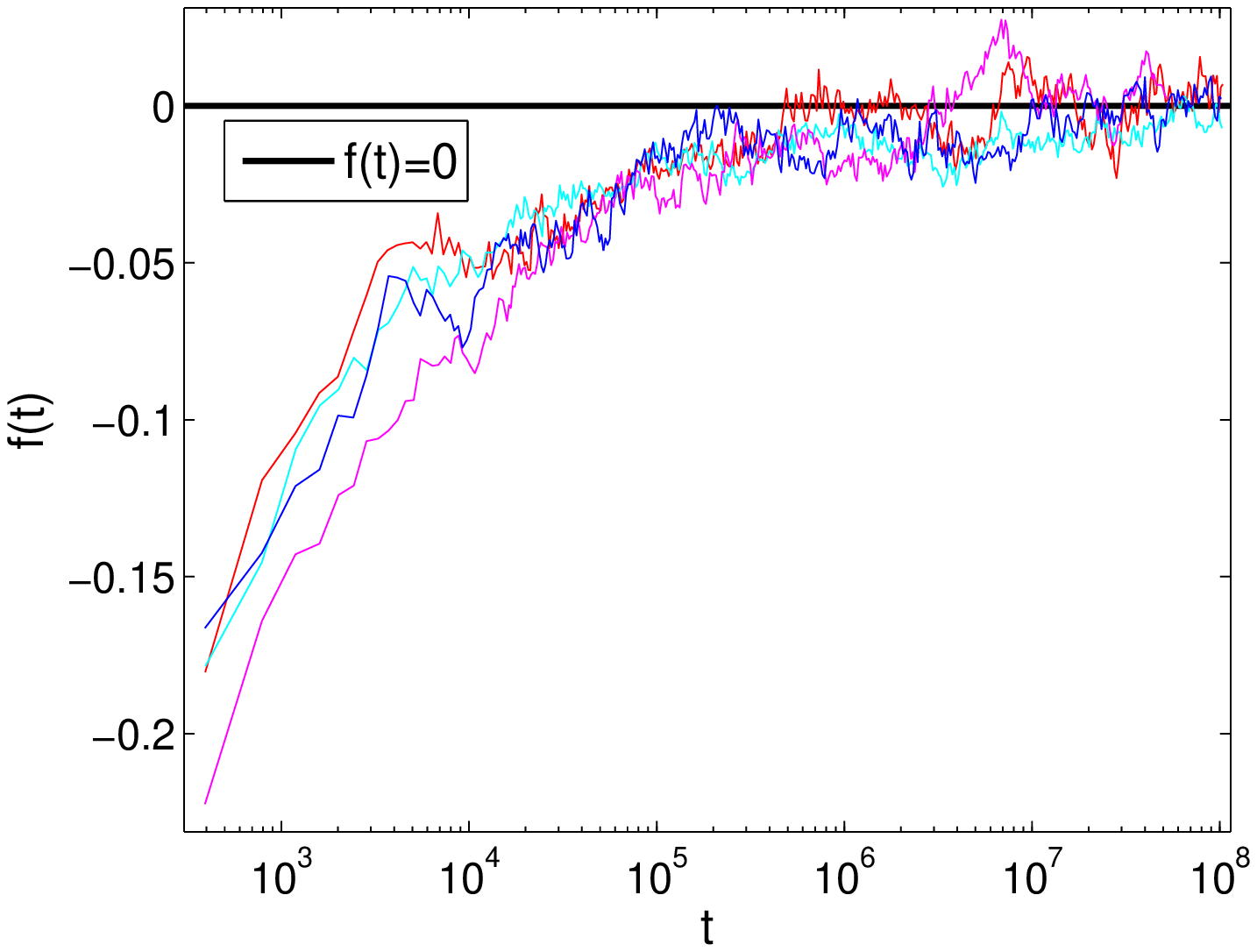}}
%\hspace {1in} \subfigure[High density]
{\includegraphics[angle=0,scale=0.6]{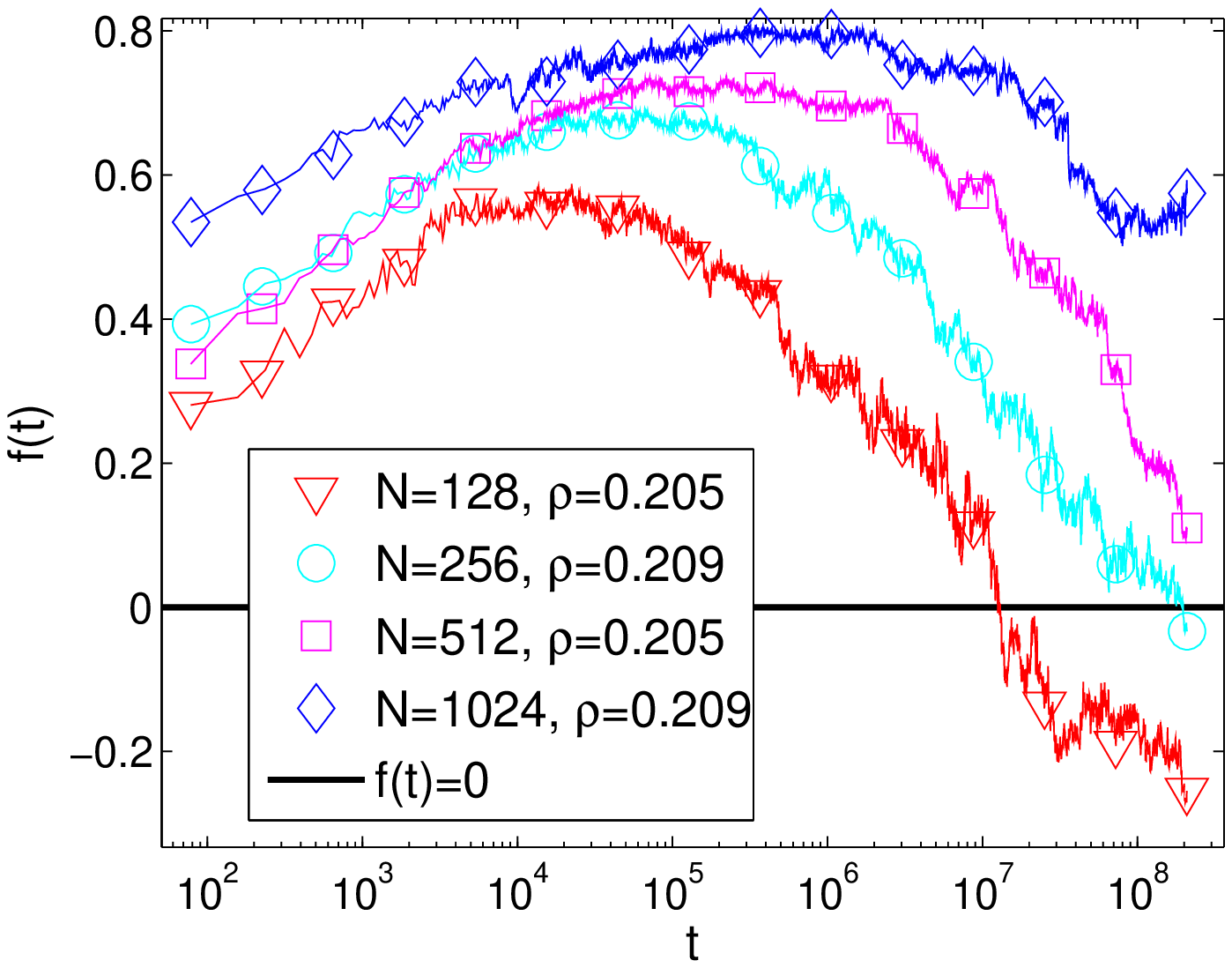}}

\end{center}
\caption{ \label{fig:f} Top: At the lowest density studied $\rho
\simeq 0.001$, the function $f$ is independent of the number of
monomers and approaches zero monotonically (i.e., the motion is
isotropic at intermediate and large time scales). Bottom: At the
highest densities achieved $\rho \simeq 0.20-0.21$, the function $f$
reaches very large positive values before decreasing again towards
zero. For short polymers $f(t)$ goes down to negative values at
large times. Here we ignore the causes of this observation because
as explained in the text, $f(t)$ is a good measure of the anisotropy
of the motion only up to certain time scales.}
\end{figure}

Let us define $f$ as a measure of the anisotropy
of the motion with respect to the longitudinal and transverse
directions,
\begin{equation}\label{f}
    f(t)=\frac{c_t(t)-c_n(t)}{c(t)}.
\end{equation}
We have $c_t(t)/c(t)=(1+f(t)) / 2$ and
$c_n(t)/c(t)=(1-f(t)) / 2$, where $f$ ranges from $-1$ to $+1$. A
large value of $f$ clearly indicates a motion primarily due to
reptation. As shown in semi-logarithmic scale in Fig.~\ref{fig:f}, $f$
reaches a maximum value of $0.8$ at short time scales, demonstrating that
reptation-like motion is the primary contributor to monomer
displacement. We also observe that the maximum of $f$ increases as
the number of monomers $N$ is increased, consistent with the fact that
the width of the plateau becomes monotonically larger and larger with $N$.
%
%
%
%
%%%%%%%%%%%%%%%%%%%%%%%%%%%%%%%%%%%%%%%%%%%%%%%%%%%%%%%%%%%%%%%%%%%%%%%%%%%%%%

\section{\label{sec: conclusions}
Conclusions
        }
As the size of the confining box around a polymer is reduced, the
monomer density makes it increasingly difficult for the polymer to
move.  However, the effect on the polymer movement is not isotropic.
The transverse fluctuations are strongly suppressed due to the
proximity of monomers that may be greatly separated along the chain
backbone.  This is in contrast to motion parallel to the chain
backbone where, due to the connectivity constraint, the monomer
density is very similar for the confined and unconfined chains.
While longitudinal motion is sub-dominant in the free chain, it is
the primary mode of monomer diffusion when the density becomes high
enough to suppress the transverse fluctuations.

The emergence of motion parallel to the chain backbone as the
dominant mode of diffusion is similar to what occurs in a polymer
solution when the density is increased to form a melt. However, the
longitudinal diffusion observed here differs from the classic
reptation picture in that the motion is not necessarily initiated at
the chain ends but it can also be triggered by fingering events. The
prevalence of fingering reptation over end-initiated reptation is
due to three factors. First, the two-dimensional nature of our
system imposes topological constraints that severely limit the
mobility of the chain ends. Second, a single confined chain has only
two end-points, while the number of fingers it can have grows with
the system size. Third, the compact configuration due to the
confinement forces the creation of more fingering structures
relative to the extended polymer structures found in melts.

The peculiarities of the dynamics of a single chain in extreme
confinement (high density limit) leads to an interesting effect:
monomers can diffuse through large distances comparable to the box
size within time scales for which the overall shape of the polymer
is, nevertheless, largely preserved. While monomer displacement
exhibits a smooth power law behavior in time at all densities, the
tangent-tangent correlation function develops a secondary decay at
high densities. This decay takes place at longer time scales for
older systems, suggestive of aging phenomena. We thus find
glass-like behavior in the overall geometry concurrently with
non-glassy monomer motion. Despite significant persistence of
geometry, monomer displacement can reach large values relative to
its saturation value over the same time scales because local
rearrangements cause monomers to flow even in parts of the system
where no reshaping is taking place.

The two dimensional lattice model presented here is a largely
simplified one.  However, we believe that this model yields
considerable insight into the generic properties of confined
polymers. Namely, reptational or longitudinal motion is identified
as the primary mechanism for motion at high densities and extreme
confinement is found to primarily suppress changes in the overall
geometry of the polymer rather than the monomer motion.
%
%
%%%%%%%%%%%%%%%%%%%%%%%%%%%%%%%%%%%%%%%%%%%%%%%%%%%%%%%%%%%%%%%%%%%%%%%%%%%%%%

\section*{
Acknowledgments
         }
The simulations were carried out on Boston University supercomputer
facilities (SCV). We thank B. Chakraborty, J. Kondev, D. Reichman
and F. Ritort for useful discussions. This work is supported in part
by the NSF Grant DMR-0403997 (AR, CC, CC, and JS) and by EPSRC Grant
No. GR/R83712/01 (C.~Castelnovo).
%
%
%%%%%%%%%%%%%%%%%%%%%%%%%%%%%%%%%%%%%%%%%%%%%%%%%%%%%%%%%%%%%%%%%%%%%%%%%%%%%%

\section*{
Appendix: Closed chains
         }
Closed chains can be studied using the same correlation functions.
The only subtlety with closed chains is the existence of a
non-trivial background in finite systems. The background has to do
with the topology of closed loops and must be subtracted from the
tangent-tangent correlation function. Suppose that the monomers in
the chain are initially indexed clockwise or anti-clockwise. The
dynamics cannot change the chirality of the loop in two dimensions.
Therefore, all the outer segments of the polymer running parallel to
the walls of the box have correlated tangent fields. Using an
ensemble with random chirality does not remove the problem because
each realization, whether clockwise or anti-clockwise, would
contribute a positive value to the correlation function.
\begin{figure}[ht]
\vspace{0.6 cm}
\begin{center}
\includegraphics[angle=0,scale=0.4]{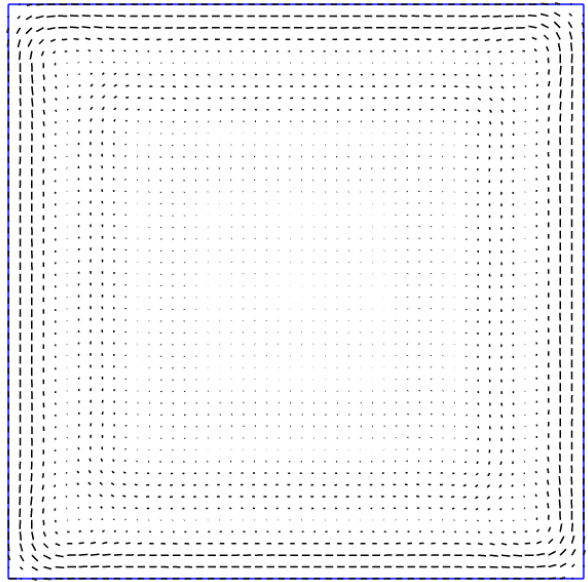}
\end{center}
\caption{
\label{fig:background}
The background tangent field for $N=256$ monomers in a box of size $L=49$.
The relative scale between the field vectors reflects the actual values
of the tangent-tangent field (an overall scale factor has been introduced
to enhance visibility).
Notice that the average field at the boundary does not vanish.
}
\end{figure}
\begin{figure}[ht]
\vspace{0.6 cm}
\begin{center}
\includegraphics[angle=0,scale=0.6]{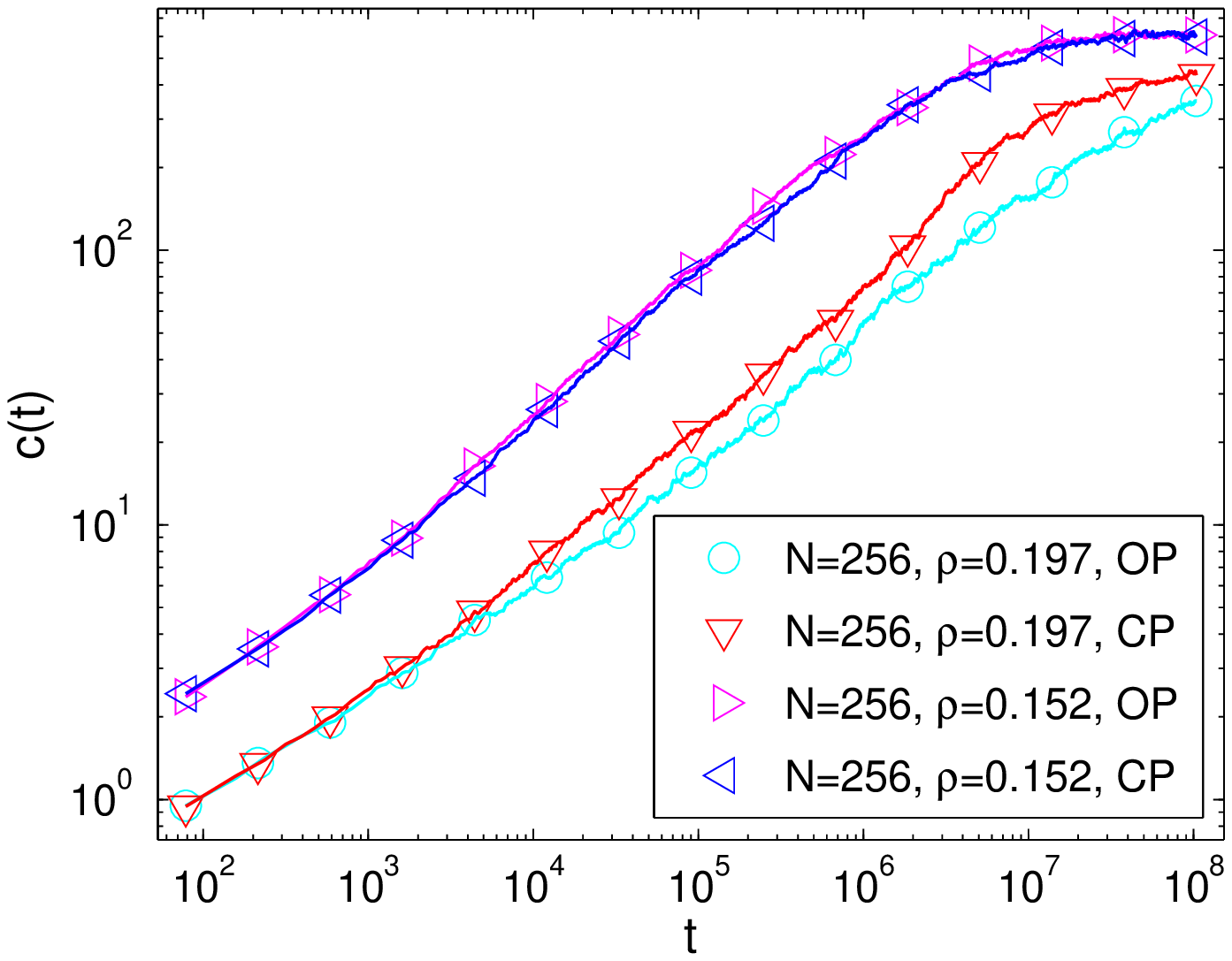}
\end{center}
\caption{ \label{fig:msdloop} The mean square displacements of
closed polymers (CP) and open polymers (OP) are identical at low and
intermediate densities. At high densities closed polymers seem to
reach the saturation value faster.}
\end{figure}
One can correct for this effect as follows.
For an ensemble with a given chirality, let us
call the equilibrium tangent-field background
$\vec{s}_{\textrm{ave}}(\vec{x})$. We can then modify the tangent-tangent
correlation function by subtracting this background field.
\begin{equation}
\label{eq: tantanloop}
c_{s, \textrm{loop}}(t,t_w)
=
\langle
  \frac{1}{N}
  \sum_{\textrm{all}\,\vec{x}}
  \left[\vphantom{\sum}
    \vec{s}_{\vec{x}}(t+t_w)
    -
    \vec{s}_{\textrm{ave}}(\vec{x})
  \right]
  \cdot
  \left[\vphantom{\sum}
    \vec{s}_{\vec{x}}(t_w)
    -
    \vec{s}_{\textrm{ave}}(\vec{x})
  \right]
\rangle.
\end{equation}
The equilibrium background can be obtained at low densities via Monte
Carlo simulations, using realizations with the same chirality and
averaging over time and ensemble. This approach becomes less and less
reliable as the density increases and glassy behavior arises, because
each realization is essentially stuck in a small region of
configuration space over the measurement time scales.
Fig.~\ref{fig:background} shows the background tangent field at an
intermediate density $\rho=0.1$. The modified tangent-tangent
correlation function~(\ref{eq: tantanloop}) and the mean square
displacement were measured for a closed loop of $N=256$ monomers and
no qualitative difference was observed in comparison to open
chains. Also $f(t)$ behaves similarly in the two cases, reaching high values
at high densities for closed chains as well as open chains.
As shown in Fig.~\ref{fig:msdloop}, the mean square
displacement for closed chains at high densities reaches its
saturation value faster than for open chains, whereas at low and
intermediate densities they are identical.
%
%
%%%%%%%%%%%%%%%%%%%%%%%%%%%%%%%%%%%%%%%%%%%%%%%%%%%%%%%%%%%%%%%%%%%%%%%%%%%%%%

\end{document}